\documentstyle[preprint,aps]{revtex}

\begin{document}

\thispagestyle{empty}

\rightline{KAIST-CHEP-94/S3}
\rightline{hep-th/9410182}

\begin{center}
\vspace{36pt}
{\large \bf Low Energy Description of Fermion Pairs in Topologically
  Massive QED$_{2+1}$ with $N$ Flavours}
\end{center}

\vspace{36pt}

\begin{center}
Jae Kwan Kim and Hyeonjoon Shin\footnote{e-mail address: {\tt
hshin@chep6.kaist.ac.kr }}
\vspace{20pt}

{\it Department of Physics,  \\
Korea Advanced Institute of Science and Technology, \\
Taejon 305-701, Korea } \\
\end{center}

\vfill

\begin{center}
{\bf ABSTRACT}
\end{center}
In topologically massive QED$_{2+1}$ with $N$ flavours, there is
the possibility that two equal-charged fermions can form a bound
state pair in either s-wave or p-wave. We are concerned about the
s-wave pairs and obtain the low energy effective action describing
them. It is shown that the fermion pairs behave like doubly charged
spin-1 bosons and, when they condense, the gauge field aquires the
longitudinal mass. The approximate $SU(2)$ symmetry due to the
similarity between the fermion pairs and the gauge field is
discussed.

\newpage
The topologically massive QED in (2+1) dimensions \cite{schon81}
is the usual QED with a Chern-Simons (CS) term and its Lagrangian is of
the form as follows:
\begin{equation}
\label{tmgt}
{\cal L}_{\rm TMQ} = -\frac{1}{4}F_{\mu\nu}F^{\mu\nu} -
	   \frac{\kappa}{4\pi} \epsilon^{\mu\nu\lambda} A_\mu
	   \partial_\nu A_\lambda - e A_\mu j^\mu + {\cal L}_m
\end{equation}
where $j^\mu$ is the matter current and ${\cal L}_m$ the kinetic term
of matter. One of the features of this model is that, due to
the structure of the CS term, a charged particle at rest
is a source of not only an electric field but
also a magnetic field. Based on this, it has been shown in ref.
\cite{kog89} that the magnetic moment leads to a {\it short-range}
interaction between particles, and the particle dynamics is drastically
changed. In the subsequent works \cite{ste91,kog91}, it has also
been shown that bosonic as well as fermionic matter can have its
magnetic moment.

For the fermionic particles, there appears an interesting possibility
that the interaction due to the magnetic moment may be attractive even
between equally charged fermions and hence it is expected that bound
states appear under certain circumstances. Some authors have investigated
on this possibility and argued that, although the exact analytical
solution has not yet been given, it is indeed possible that such
bound states appear \cite{kog89,gro90,gir92,dob93}.

The strength and the form of the interaction between two equally charged
fermions mediated by topologically massive photon are obtained by
calculating the amplitude ${\cal M}$ of M\"{o}ller scattering.
In refs.~\cite{kog89,kog91,gro90,gir92,dob93}, the tree level calculations
of ${\cal M}$ have been done in the nonrelativistic limit as follows:
i) For the scattering of fermions with different types in an
antisymmetric manner, the amplitude, say ${\cal M}_s$, gives a pure
s-wave interaction ${\cal M}_s = -(e^2/m_T^2) (m_T/m -1)$ where
$m_T~(= \kappa /2 \pi)$ is the topological (transverse) mass of the
gauge field and $m$ the fermion mass. ii) For the scattering of identical
fermions, the amplitude, say ${\cal M}_p$, gives a pure p-wave interaction
${\cal M}_p = -2 (e^2/m_T^4) (m_T/m-1) {\bf p} \cdot {\bf p}'$ where
${\bf p}~({\bf p}')$ is the momentum of the incoming (outgoing)
fermions in the center of mass frame.
These imply that, for $m_T/m > 1$, s- and p-wave fermion-fermion bound
states may exist. Now this argument may be doubtful since it has been
given in the tree level approximation. However, since the tree level
results remain unchanged at least at one-loop order \cite{sza93}, the
arguments for the bound states still remain.

Besides on the existence of the fermion-fermion bound state pair itself,
one may now consider an interesting problem of what behavior the pairs
show up when they form. In ref.\cite{kog89}, this problem was considered
for the case of identical fermions, where only the p-wave pairing is
possible, and the possible vacuum instability was investigated by solving
the gap equation. In this paper, we are concerned about the problem in
the case where the pairing occurs in the s-wave and treat it by
following the procedure of obtaining the Landau-Ginzburg description
of Cooper pairs from the microscopic BCS theory \cite{sak}.

The Lagrangian (\ref{tmgt}) itself would not be considered in our work.
Instead of it, by mimicking the BCS theory, we consider the Lagrangian
where the short-range interaction term of fermions due to photon
exchange is explicitly included. The form of the interaction term can
be inferred from ref.\cite{mac94} where the similar situation with ours
was presented. In there, the interaction being responsible for the pairing
of two equally charged fermions was the four-fermion one and was represented
by using the doubly charged bilinear in the fermion. The bilinear was
the scalar type and given by introducing the charge conjugated fermion
field, $\psi^c =C \bar{\psi}^T$, $C$ being the charge conjugation matrix.
In our case, the interaction term must be the vector type since the gauge
field leads to the vector type interaction as can be shown from
(\ref{tmgt}). And the fact that the s-wave pairing is that of fermions
with different types in an antisymmetric manner leads us to introduce
at least two types of fermions, which we label them as A type $(\psi_A)$
and B type $(\psi_B)$. Then the relevant interaction term can be written
down as follows:
\begin{equation}
\label{ffint}
g(\bar{\psi}\gamma_\mu\tau^T \psi^c) (\bar{\psi}^c\gamma^\mu\tau\psi)~,
\end{equation}
where $g$ is the coupling constant and $\psi^T = ( \psi_A , \psi_B)$
is the four component spinor ($\psi_A$ and $\psi_B$ are two component
spinors). Here the gamma matrices and $\tau$ matrix is given by
\begin{equation}
\label{4gam}
\gamma^0 = \left( \begin{array}{cc}
		    \sigma^3 & 0 \\ 0 & \sigma^3
                  \end{array}
           \right) ~, ~~~
\gamma^j = \left( \begin{array}{cc}
		    i \sigma^j & 0 \\ 0 & i \sigma^j
                  \end{array}
	   \right)~, ~~~
\tau = \left( \begin{array}{cc} 0 & I \\ -I & 0 \end{array}
       \right)~,
\end{equation}
\[ (j=1,2) \]
where $\sigma^i~(i=1,2,3)$ are the usual Pauli matrices and $I$ the
unit $2 \times 2$ matrix.  The matrix $\tau$ plays the role of
mixing A and B type fermions antisymmetrically. With these
representations, the relations\footnote{Our conventions are
$g^{\mu\nu} = {\rm diag} (1,-1,-1)$, $\epsilon^{012} = +1$ and
$\epsilon^{ij}=\epsilon^{0ij}$.}
\begin{eqnarray}
& \{ \gamma^\mu, \gamma^\nu \} = 2 g^{\mu\nu} ~,& \nonumber \\
& \gamma^\mu \gamma^\nu = g^{\mu\nu} - i \epsilon^{\mu\nu\rho}
  \gamma_\rho ~,& \label{grel} \\
& [ \gamma^\mu, \tau ] = 0  & \nonumber
\end{eqnarray}
are satisfied and the charge conjugation matrix $C$ becomes
\begin{equation}
C = \gamma_2~.
\end{equation}
Now, by using the on-shell positive energy spinor
\cite{sza93}, one can easily check that, in the nonrelativistic
limit, the interaction (\ref{ffint}) gives just the s-wave result
of ${\cal M}_s$ and
\begin{equation}
\label{ccr}
g \propto \frac{m_T}{m}-1 ~,
\end{equation}
which means that the interaction is attractive if $g>0$. Thus it is
concluded that the interaction (\ref{ffint}) describes correctly
the interaction of fermions in the theory (\ref{tmgt}) at least
in the low energy region. By the way, the coupling constant $g$
has negative mass dimension, $[g]=-1$, and hence the four-fermion
interaction (\ref{ffint}) is not renormalizable under the
conventional perturbation method. However, such interaction
turns into renormalizable one in the large $N$ perturbation
theory where $N$ is the number of fermion flavours \cite{ros89}.
This naturally leads us to introduce fermion flavours. Then the
Lagrangian we consider is given by
\begin{equation}
\label{ffsys}
{\cal L} = \bar{\psi}_i(i{\not\!\partial} -e{\not\!\!A} -m)\psi_i
	   - \frac{g}{N} (\bar{\psi}_i \gamma_\mu\tau^T \psi^c_i)
			 (\bar{\psi}^c_j\gamma^\mu\tau\psi_j)
\end{equation}
where flavour indices $i$ and $j$ range from $1$ to $N$. The
gauge field $A_\mu$ is purely external and the Maxwell-CS
kinetic term for it is omitted here. As an important symmetry,
this Lagrangian is invariant under the $U(1)$ local gauge
transformations: $\psi \rightarrow e^{i\Lambda}\psi$ and
$A_\mu \rightarrow A_\mu - (1/e)\partial_\mu \Lambda$.

The four-fermion term in (\ref{ffsys}) may be linearized by
introducing an auxiliary complex vector field $\chi_\mu$.
The linearized form of (\ref{ffsys}) is then
\begin{equation}
\label{ffsys2}
{\cal L} = \bar{\psi}_i(i{\not\!\partial}-e{\not\!\!A}-m)\psi_i
	   - \chi^*_\mu \bar{\psi}^c_i\gamma^\mu\tau\psi_i
	   - \chi_\mu \bar{\psi}_i \gamma^\mu\tau^T \psi^c_i
	   + \frac{N}{g} \chi^*_\mu \chi^\mu ~.
\end{equation}
In order to maintain the gauge invariance of ${\cal L}$, the
vector field should transform as $\chi_\mu \rightarrow
e^{i 2 \Lambda}\chi_\mu$, which means that the charge of the
vector field is $2e$. By analogy with the BCS theory, the
vector field $\chi_\mu$ in (\ref{ffsys2}) is interpreted as the
condensate and describes the fermion-fermion pair, which may
be viewed from the equation of motion for the field,
$\chi_\mu = (g/N) \bar{\psi}^c \gamma_\mu \tau \psi$.
In what follows, we calculate the low energy effective action
for the vector and the gauge fields, after the investigation
of the phase structure according to the coupling constant
$g$. All the formulations are performed to leading order in
$1/N$ and the flavour indices are omitted from now on. As a
remark, it should be emphasized that only the low energy
effective action is meaningful since the four-fermion term in
(\ref{ffsys}) reflects only the low energy aspect of the
interaction between fermions in the system (\ref{tmgt}).

The Lagrangian (\ref{ffsys2}) is bilinear in the fermion field,
and may be integrated out for the field. Although it is so,
some care is needed because of the presence of the charge
conjugated field, $\psi^c$. We first recall the invariance of
$\bar{\psi}(i{\not\!\partial}-e{\not\!\!A}-m)\psi$ under the
charge conjugation;
\begin{equation}
\label{ccinv}
\bar{\psi}^c(i{\not\!\partial}-e {\not\!\!A}^c -m)\psi^c =
\bar{\psi}(i{\not\!\partial}-e{\not\!\!A}-m)\psi
\end{equation}
where the charge conjugated gauge field $A^c_\mu$ is given by
$-A_\mu$ as usual. By using (\ref{ccinv}) and introducing new
field variable $\theta$,
\begin{equation}
\label{tfld}
\theta = \frac{1}{\sqrt{2}} \left( \begin{array}{c}
		                     \psi \\ \psi^c
                            \end{array} \right)~,
\end{equation}
we can rewrite the Lagrangian (\ref{ffsys2}) as follows:
\begin{equation}
\label{sys}
{\cal L} = \bar{\theta} K \theta +
	   \frac{N}{g} \chi^*_\mu \chi^\mu
\end{equation}
where $K$ is a matrix and defined by
\begin{equation}
\label{kmat}
K =  \left( \begin{array}{cc}
	       i{\not\!\partial}-e{\not\!\!A}-m &
	       -2 \chi_\mu \gamma^\mu \tau^T \\
	       -2 \chi^*_\mu \gamma^\mu \tau &
	       i{\not\!\partial}+e{\not\!\!A}-m
            \end{array}
     \right)~.
\end{equation}
This Lagrangian is clearly bilinear in the $\theta$
fields.

Now the effective action for the vector and gauge fields
may be attained by integrating over the $\theta$ fields:
\begin{equation}
\exp i S_{\rm eff} [ \chi, \chi^*, A ]
  = \int D \bar{\theta} D \theta \exp \left( i
                   \int d^3x {\cal L} \right)~.
\end{equation}
The formal expression of $S_{\rm eff}$ is
\begin{equation}
\label{eff}
S_{\rm eff} =  \frac{N}{g} \int d^3 x \chi^*_\mu \chi^\mu
	      - i \frac{N}{2} \ln \det K
\end{equation}
where $\det$ is the functional determinant. It should be
noted here that the factor $1/2$ is multiplied to the
second term of the r.h.s. of eq.~(\ref{eff}) since the
$\theta$ integration gives double the result obtained
from the original $\psi$ integration as can be known
from the definition of $\theta$ (\ref{tfld}). $\det K$ in
(\ref{eff}) may not be directly evaluated because of the
matrix nature of $K$. To make it more tractable, we
decompose $K$ by using the following identity:
\begin{equation}
\left( \begin{array}{cc}
         A & B \\ C & D
       \end{array}
\right) =
\left( \begin{array}{cc}
       A & 0 \\ C & 1
       \end{array}
\right)
\left( \begin{array}{cc}
       1 & A^{-1}B \\ 0 & D-CA^{-1}B
       \end{array}
\right) ~.
\end{equation}
We then obtain, after a few manipulations,
\begin{equation}
\label{detk}
\det K = {\det}^2 (i{\not\!\partial}-e{\not\!\!A}-m)
           \det \left( 1 - 4
	     \frac{1}{i{\not\!\partial}+e{\not\!\!A}-m}
	     \chi^* \cdot \gamma
	     \frac{1}{i{\not\!\partial}-e{\not\!\!A}-m}
	     \chi \cdot \gamma \right)~,
\end{equation}
where we use the fact that
$\det (i{\not\!\partial}+e{\not\!\!A}-m)$ is equal to
$\det (i{\not\!\partial}-e{\not\!\!A}-m)$
due to Furry's theorem and remove the $\tau$ matrices by using
eq.~(\ref{grel}). By substituting this into
eq.~(\ref{eff}), the effective action becomes
\begin{equation}
\label{teff}
S_{\rm eff} = S_A + S_{\chi A}
\end{equation}
where
\begin{equation}
\label{aeff}
S_A = -i N {\rm Tr} \ln (i{\not\!\partial}-e{\not\!\!A}-m)~,
\end{equation}
\begin{equation}
\label{ceff}
S_{\chi A} = \frac{N}{g} \int d^3x \chi^*_\mu \chi^\mu
         - i \frac{N}{2} {\rm Tr} \ln \left( 1 - 4
	     \frac{1}{i{\not\!\partial}+e{\not\!\!A}-m}
	     \chi^* \cdot \gamma
	     \frac{1}{i{\not\!\partial}-e{\not\!\!A}-m}
	     \chi \cdot \gamma \right)~.
\end{equation}
Here, we used the identity $ \ln \det X = {\rm Tr} \ln X$ where Tr
is the functional trace. If we set the field $\chi_\mu$ equal to zero,
the effective action reduces to the usual fermion determinant,
$S_A$, as it should be.

Before we calculate the effective action, we first
concentrate on the phase structure in terms of the coupling
constant $g$. The phase structure is obtained by
investigating the saddle point or gap equation involving the
vacuum expectation value $v$ of the field $\chi_\mu$. Here,
$v$ is taken as the Lorentz invariant scalar, and given by
\begin{equation}
v = \langle | \chi | \rangle ~,
\end{equation}
with $|\chi| = (\chi^*_\mu \chi^\mu)^{1/2}$.
For the gap equation, we need to calculate the effective
potential $V_{\rm eff}$, which is defined as
$\int d^3x V_{\rm eff} = - S_{\rm eff}$ where the value
of $\chi_\mu$ is constant and the gauge field is turned off:
\begin{equation}
\label{effp}
V_{\rm eff} = - \frac{N}{g} \chi^*_\mu \chi^\mu
         + i \frac{N}{2} \int^\Lambda \frac{d^3 p}{(2\pi)^3}
	   {\rm tr} \ln \left( 1 - 4
	                  \frac{1}{ \not\!p -m} \gamma^\mu
	                  \frac{1}{ \not\!p -m} \gamma^\nu
	                  \chi^*_\mu \chi_\nu
	        	 \right) ~,
\end{equation}
with $\Lambda$ the spatial momentum cutoff and ${\rm tr}$
the trace of gamma matrices. This potential, however, is not
computed easily because the structure of the gamma matrices
in the logarithmic function of eq.~(\ref{effp}) is not
simplified. Here, we restrict ourselves to the case where the
value of $\chi_\mu$ is close to zero. In this case, $V_{\rm eff}$
may be expanded in powers of $\chi_\mu$ and hence the gap equation
may be obtained approximately. If we expand $V_{\rm eff}$ up
to the first nontrivial order, then the gap equation
becomes, after a straightforward calculation,
\begin{eqnarray}
0 &=& \frac{1}{N} \frac{\delta V_{\rm eff}}{\delta v}
	 \nonumber \\
  &\simeq& 2 v \left( - \frac{1}{g} + \frac{2}{3\pi}
		      \sqrt{\Lambda^2+m^2} +
		      \frac{8 v^2}{9\pi |m|}
               \right) ~. \label{geq}
\end{eqnarray}
The solution of this equation with nonvanishing $v~(>0)$
exists only when $g$ is smaller than the critical coupling
$g^{-1}_c \equiv (2/3\pi) \sqrt{\Lambda^2+m^2}$. In that
region, $v$ is given by
\begin{equation}
\label{vev}
v^2 \simeq \frac{9 \pi |m|}{8} \left( \frac{1}{g} - \frac{1}{g_c} \right)
 ~, ~~ 0<g<g_c ~.
\end{equation}
This clearly shows that the condensation of fermion-fermion
pairs occurs in the weak coupling phase. This result is
the reasonable one if we recall the previous discussion on
the nonrelativistic limit of the interaction (\ref{ffint}), in particular
eq.~(\ref{ccr}) ($g>0$ corresponds to the attraction between fermions).
In refs.~\cite{kog89,dob93}, it has been argued that, if $m_T$ becomes
very large so that $m_T \gg m$, the fermion-fermion bound states would
not form in the theory (\ref{tmgt}). This is further supported by
the weak coupling result (\ref{vev}), since, according to eq.~(\ref{ccr}),
the region of $m_T \gg m$ corresponds to the strong coupling phase where
the condensation does not appear.
Now we would like to note that eq.~(\ref{vev}) is valid only in the region
near the critical point; if $g$ is far from $g_c$,
eq.~(\ref{vev}) cannot be trusted.

We now turn our interest to the effective action $S_{\rm eff}$. For its
evaluation, we take the derivative expansion method \cite{fra85}, which
is well suited for obtaining the low energy effective action. The fermion
determinant $S_A$ would not be evaluated, since it is well known that its
evaluation leads just to the CS action in the low energy limit
\cite{red84}. In particular, the evaluation using the derivative expansion
was also given in ref. \cite{bab87} where, of course, the same result was
reproduced. Thus we would only quote the result which is given by
\begin{equation}
\label{sa}
\frac{1}{N} S_A \simeq - \frac{e^2}{4\pi} \frac{m}{|m|} \int d^3 x
  \epsilon^{\mu\nu\lambda} A_\mu \partial_\nu A_\lambda ~.
\end{equation}
If we recover the original CS action for the gauge field which was
omitted in the Lagrangian (\ref{ffsys}), this action leads just up to the
renormalization of the CS coefficient $\kappa$, and the renormalized
coefficient, $\kappa_r$, becomes $\kappa_r = \kappa + \frac{Ne^2m}{|m|}$.
We evaluate the $S_{\chi A}$ up to quadratic order in the field $\chi_\mu$
in the low energy limit and near the critical point $g_c$, where the
vacuum expectation value $v$ is very close to zero and hence the
lowest order contribution in $v$ is dominant. The primary
contribution comes from the term containing single derivative,
$\chi^* \partial \chi$. Since the vector field $\chi_\mu$ transforms
under the gauge transformation, this term is not gauge invariant. To make
it gauge invariant, we must consider the coupling term with the gauge
field. The coupling terms may be obtained by expanding the denominators of
eq. (\ref{ceff}) in powers of the gauge field. The first one, which is
needed here, is of the form of $A \chi^* \chi$. The non derivative
quadratic term is $\chi^* \chi$ and may be inferred from the gap equation
(\ref{geq}). If we assemble the calculations for the three terms until
now, we get the expression of $S_{\chi A}$ as follows:
\begin{equation}
\label{sca}
\frac{1}{N} S_{\chi A} \simeq \frac{m}{\pi |m|} \int d^3 x \chi^*_\mu
\left[ -\epsilon^{\mu\nu\lambda} ( \partial_\nu + 2i e A_\nu )
       + g^{\mu\lambda} M
\right] \chi_\lambda
\end{equation}
with $M \equiv\pi\frac{m}{|m|}\left( \frac{1}{g}-\frac{1}{g_c} \right)$.
This gives the Landau-Ginzburg like description for the s-wave
fermion-fermion pairs existing in system (\ref{tmgt}) in the low energy
and the critical region. If we recall the transformation of the field
$\chi_\mu$ under the gauge transformation, this action is clearly gauge
invariant.

The behavior of the field $\chi_\mu$ may now be read off from the
effective action (\ref{sca}). Specifically, we look at the equation of
motion for the field derived as follows:
\begin{equation}
\label{ceq}
[ - \epsilon^{\mu\nu\lambda} (\partial_\nu + 2ieA_\nu ) + g^{\mu \lambda} M
] \chi_\lambda = 0~.
\end{equation}
If we ignore the gauge field for a moment, this is just the equation of motion
for the spin 1 field following from the representation theory of the
Poincar\'{e} algebra in 2+1 dimensions \cite{jac91,che94}. Thus the quanta of
the field $\chi_\mu$ behave like spin 1 vector bosons with the charge of
amount $2e$. However, it should be noted that this is true only in the case
of the weak coupling phase. At the critical point, $g=g_c$, the mass term
for the field $\chi_\mu$ in (\ref{sca}) disappears and hence the field
$\chi_\mu$ becomes massless. Since the massless field is spinless in
2+1 dimensions \cite{schon81}, the field $\chi_\mu$ is so in this case, and
it is explicitly distinguished from that in the weak coupling case.
Above the critical point, that is, in the strong coupling phase, $g>g_c$,
there is no condensation of $\chi_\mu$. This implies that the
fermion-fermion pairs do not form and hence it is meaningless to consider
the behavior of the field $\chi_\mu$ describing them.

It is not so difficult to understand the fact that $\chi_\mu$ is the spin
1 field in the weak coupling phase. In 2+1 dimensions, the massive
Dirac fermion has only one spin degree of freedom; the spin of it can take
only one of two possible values $\pm \frac{1}{2}$ \cite{schon81,boy86}. If
we specify $m>0$, the spin of the fermion (antifermion) is $+ \frac{1}{2} ~
( - \frac{1}{2})$. Thus the spin of the fermion-fermion pair is
simply given by adding spin values of two constituent fermions, and becomes
$+1$. This contrasts with that of Cooper pairs in the usual BCS theory: the
Cooper pair is constituted by two electrons with spin up and down
respectively, and is described by spin zero scalar field.

Now we would like to address the question whether or not the Meissner effect
takes place in the weak coupling phase. This may be answered by
investigating the appearance of the longitudinal mass of the gauge field.
The related term in the derivative expansion of (\ref{ceff}) is
$A^2 \chi^* \chi$. The direct calculation for it yields the result
as follows:
\begin{equation}
N \frac{8e^2}{\pi |m|} \int d^3 x A_\mu A^\mu \chi^*_\nu \chi^\nu ~,
\end{equation}
which is to be considered as one term in the effective action. This clearly
shows that, if condensation occurs, the gauge field aquires the
longitudinal mass depending on the value $16 N e^2 v^2 / \pi |m|$ or
$18 N e^2 m M / \pi |m|$. The longitudinal mass is the origin of the
screening of external magnetic field, i.e. Meissner effect. Thus it may
be concluded that the Meissner effect occurs in the weak coupling phase.

There is the similarity between the field $\chi_\mu$ and the gauge field
that they are all massive vector fields in the weak coupling phase.
This causes us to expect the other symmetry structure in addition to
the $U(1)$ gauge symmetry. In ref. \cite{dob93}, it was speculated that
fermion-fermion and antifermion-antifermion pairs and the gauge field
would form an $SU(2)$ isospin triplet state in a certain situation,
but without any detail analysis. We now give an attempt for the
realization of the symmetry with the low energy effective Lagrangian.
Note that the symmetry is approximate one unlike the $U(1)$ gauge
symmetry, since it would be valid only in the low energy limit.
If we choose $m>0$ for certainty, the effective Lagrangian,
$\int d^3x {\cal L}_{\rm eff} = S_{\rm eff}$, is written as
\begin{eqnarray}
{\cal L}_{\rm eff} &=& \frac{N}{\pi} \chi^*_\mu
 \left[ - \epsilon^{\mu\nu\lambda} ( \partial_\nu + 2ieA_\nu)
        + g^{\mu \lambda} M \right] \chi_\lambda \nonumber \\
  & & - \frac{\kappa_r}{4 \pi} \epsilon^{\mu\nu\lambda} A_\mu \partial_\nu
       A_\lambda + \frac{N}{\pi} 9e^2 M A_\mu A^\mu + O(\partial^2, v^4)
       \label{lag}
\end{eqnarray}
where the origianl CS term omitted in (\ref{ffsys}) is recovered.
Although the Lagrangian seems to be gauge noninvariant, the gauge
symmetry of it is guaranteed if the higher derivative terms and the
higher polynomials in fields are considered. We first set
\begin{equation}
\chi_\mu = B^1_\mu + i B^2_\mu~, ~~~
A_\mu = 2 \sqrt{\frac{N}{\kappa_r}} B^3_\mu
\end{equation}
and substitute these into (\ref{lag}). Then we obtain
\begin{equation}
\label{lag2}
{\cal L}_{\rm eff} = - \frac{N}{\pi} \epsilon^{\mu\nu\lambda}
     \left( B^a_\mu \partial_\nu B^a_\lambda +
	    \frac{8 e \sqrt{N/\kappa_r}}{3} \epsilon^{abc} B^a_\mu
	    B^b_\nu B^c_\lambda
     \right) + \frac{N}{\pi} M^{ab} B^a_\mu B^{b \mu} +O(\partial^2,v^4)
\end{equation}
where the isospin indices $a$, $b$, and $c$ take the values of 1, 2 and
3 and the mass matrix $M^{ab}$ is given by
\begin{equation}
M^{ab} = \left( \begin{array}{ccc}
		   ~1~ & 0 & 0 \\
		   0 & ~1~ & 0 \\
		   0 & 0 & \frac{36e^2 N}{\kappa_r}
                \end{array}
         \right) M~.
\end{equation}
Now, if $\kappa_r$ takes the special value of $36e^2N$, the Lagrangin
(\ref{lag2}) is invariant under the global $SU(2)$ isospin
transformation,
\begin{equation}
B^a_\mu \rightarrow U^{ab} (\phi ) B^b_\mu ~,
\end{equation}
where the matrix $U(\phi) = \exp ( i\phi^a T^a)$ is the element of the
$SU(2)$ group in the adjoint representation with the generators
$(T^a)^{bc} = -i \epsilon^{abc}$. Thus, in this special case, the field
$\chi_\mu$, its conjugate field and the gauge field form an $SU(2)$
isospin triplet in the long wave-length limit and in the weak coupling
phase. At the critical point, the mass matrix $M^{ab}$ vanishes, and
the Lagrangian becomes just the non-Abelian CS theory which enjoys the
local $SU(2)$ symmetry. This yields an interesting conclusion that the
low energy nature of the system would become topological at the
critical point.

In summary, we have been concerned about the s-wave fermion-fermion pairs
which may appear in the topologically massive QED$_{2+1}$ with some
generalization. By considering the Lagrangian where the appropriate
four-fermion term responsible for the pairing is explicitly included,
we have given the Landau-Ginzburg like description for them in the low
energy limit and near the critical point. It has been
shown that they are described by doubly charged spin 1 vector bosons
and, when they condense, the gauge field acquires the longitudinal mass
indicating the Meissner effect. Provided that $\kappa_r = 36 e^2 N$,
it has been realized that there appears the global $SU(2)$ isospin
symmetry due to the similarity between the condensate fields and the
gauge field in the condensation phase.

\section*{Acknowledgments}
One of us (H.~S.) would like to thank Dr.~W.~T.~Kim for
his continuous interest in this work and encouragement.
This work was supported in part by Korea Science and
Engineering Foundation (KOSEF) and in part
by the Center for Theoretical Physics and Chemistry.

\end{document}